\documentclass[letterpaper]{article} 
\usepackage{aaai23}  
\usepackage{times}  
\usepackage{helvet}  
\usepackage{courier}  
\usepackage[hyphens]{url}  
\usepackage{graphicx} 
\usepackage{natbib}  
\usepackage{caption} 
\usepackage{algorithm}
\usepackage{algorithmic}

\usepackage{xspace}
\usepackage{amsmath}
\usepackage{amsfonts}
\usepackage{multirow}
\usepackage{array}
\usepackage{anyfontsize}
\usepackage{subcaption} 

\usepackage{newfloat}
\usepackage{listings}
\DeclareCaptionStyle{ruled}{labelfont=normalfont,labelsep=colon,strut=off} 
\floatstyle{ruled}
\newfloat{listing}{tb}{lst}{}
\floatname{listing}{Listing}
\title{Dynamic Multi-Behavior Sequence Modeling for Next Item Recommendation}
\author{
    Junsu Cho\textsuperscript{\rm 1},
    Dongmin Hyun\textsuperscript{\rm 2},
    Dong won Lim\textsuperscript{\rm 3}, 
    Hyeon jae Cheon\textsuperscript{\rm 3},
    Hyoung-iel Park\textsuperscript{\rm 3},
    Hwanjo Yu\textsuperscript{\rm 1}\thanks{Corresponding Author.}
}
\affiliations{
    \textsuperscript{\rm 1}Dept. of Computer Science and Engineering, POSTECH, Republic of Korea\\
    \textsuperscript{\rm 2}Institute of Artificial Intelligence, POSTECH, Republic of Korea\\
    \textsuperscript{\rm 3}GS Retail, Republic of Korea\\

    \{junsu7463, dm.hyun\}@postech.ac.kr, \{dlaehddnjs99, redhat2k, hi.park\}@gmail.com, hwanjoyu@postech.ac.kr
}

\newcommand{\proposedA}{DyMuS\xspace}
\newcommand{\proposedB}{DyMuS$^\text{+}$\xspace}

\begin{document}

\maketitle

\begin{abstract}
Sequential Recommender Systems (SRSs) aim to predict the next item that users will consume, by modeling the user interests within their item sequences. 
While most existing SRSs focus on a single type of user behavior, only a few pay attention to \textit{multi-behavior sequences}, although they are very common in real-world scenarios. 
It is challenging to effectively capture the user interests within multi-behavior sequences, because the information about user interests is entangled throughout the sequences in complex relationships. 
To this end, we first address the characteristics of multi-behavior sequences that should be considered in SRSs, and then propose novel methods for \textbf{Dy}namic \textbf{Mu}lti-behavior \textbf{S}equence modeling named \proposedA, which is a light version, and \proposedB, which is an improved version, considering the characteristics. 
\proposedA first encodes each behavior sequence independently, and then combines the encoded sequences using \textit{dynamic routing}, which dynamically integrates information required in the final result from among many candidates, based on correlations between the sequences.  
\proposedB, furthermore, applies the dynamic routing even to encoding each behavior sequence to further capture the correlations at item-level. 
Moreover, we release a new, large and up-to-date dataset for multi-behavior recommendation. 
Our experiments on \proposedA and \proposedB show their superiority and the significance of capturing the characteristics of multi-behavior sequences. 
\end{abstract}

\section{Introduction}
In the era of information overload, Recommender Systems (RSs) have played an important role in helping users to discover which items to consume from a large amount of items, by modeling the user interests.
As the user interests on items drift over time, Sequential Recommender Systems (SRSs) \cite{fmlp, proxysr, peris} which capture the sequential dynamics of user interests have shown outstanding performance. 


Although there have been many studies for SRSs, most of them consider a single type of behavior and only a few studies take into account \textit{multi-behavior sequences}. 
In real-world scenarios, however, users often take several kinds of behaviors such as click, add-to-cart, add-to-favorite, and purchase. 
Compared to the single-behavior data, the multi-behaviors of users provide diverse perspectives of user interests, which conjointly imply the context of user interests and causal relationships between the user behaviors \cite{tgt}. 
For learning such various information and effectively utilizing it to the next item recommendation, the sequence modeling for multi-behavior sequences needs to consider some unique characteristics of multi-behavior sequences.

In this paper, we first address the characteristics of multi-behavior sequences that should be considered in SRSs: 
1) The data distribution of each behavior type is \textit{imbalanced}. 
For example, users usually perform more clicks than add-to-carts or purchases. 
2) The multi-behavior sequences involve \textit{heterogeneous information} about user interests, so each can provide complementary information in predicting the users' next item. 
For example, purchase data implies the user's general preferences while recent click data indicates the category of the item the user is currently looking for. 
3) The key information involved in each type of behavior sequence and its importance is \textit{personalized} to the user.
For example, a behavior sequence may indicate the user interest on an item category, brand or price, or it may not be important, depending on the user's intention for that behavior.
Therefore, the model should be able to extract necessary information from a behavior sequence according to the user. 
4) There are \textit{correlations} between the behavior sequences. 
In other words, important information of a behavior sequence may be determined according to the information of other sequences. 
For example, if a user recently clicked on items in similar categories, it is also likely to affect on the purchase sequence.




The existing SRS models, for single-behavior or multi-behavior sequences, do not take into account the characteristics of multi-behavior sequences addressed above in depth. 
For example, a single unified sequence for the multi-behavior data is unmanageably too long to contain a sufficient amount of sparse behaviors. 
In addition, it is difficult to learn heterogeneous and personalized information from the various types of behavior sequences with some coarse-grained vector representations \cite{mbn}. 

To this end, this paper proposes two novel modeling methods that captures the characteristics of multi-behavior sequences mentioned above, named \proposedA and \proposedB. 
Our proposing methods encode each behavior sequences and combine them, rather than handling a single sequence containing all the behaviors in order to deal with the behavior imbalance problem.
\proposedA combines the information encoded from each behavior sequence using the concept of \textit{dynamic routing} \cite{capsnet}, which dynamically combines important information required in the combined result from among many candidates.
The dynamic routing in \proposedA integrates \textit{personalized information} for the user among candidate capsules \cite{capsule} encoding \textit{heterogeneous information} of multi-behavior sequences, based on the \textit{correlations} between them. 
\proposedB, further applies the dynamic routing to the modeling of each behavior sequence to consider item-level heterogeneity and personalization based on the correlations.
We also resolve the scalability issue that occurs when directly applying the dynamic routing to recommendation. 
In brief, our methods dynamically determine the important heterogeneous information of multi-behavior sequences for the user at sequence- and item-level based on the correlations between sequences, capturing all of the aforementioned characteristics of multi-behavior sequences. 

This paper releases a new dataset, \textit{GS Retail}, for Multi-Behavior Recommender System (MBRS), which is collected from a real-world e-commerce service in South Korea. 
This dataset contains larger and more up-to-date data than the existing public datasets commonly used for MBRSs. 
Our extensive experiments on existing public datasets as well as our new dataset demonstrate that our proposed methods considerably outperform various state-of-the-art baselines. 
Also, our analyses show that the ability of \proposedA and \proposedB to capture the characteristics of multi-behavior sequences presented above significantly affects the recommendation performance.



\section{Related Work}

\subsection{Sequential Recommender Systems}
SRSs aim to predict the next item a user will consume using the sequence of items consumed by the user. 
Many studies on SRS focus on how to effectively learn the user's long-term and short-term interest from the sequence. 
For example, GRU4Rec \cite{gru4rec} models the user interests by encoding the sequence using GRU \cite{gru}. 
SASRec \cite{sasrec} uses self-attention mechanism \cite{transformer} to consider the long- and short-term interest in the item sequence. 
Recently, FMLP-Rec \cite{fmlp} based on an all-MLP structure reduces the noises in the sequence using filtering algorithms. 

Although there are many studies on SRS, only a few pay attention to the multi-behavior scenario.
Compared with the SRSs based on single-behavior sequences, SRSs based on multi-behavior sequences can take advantage of the fine-grained information about user interests from the multi-behavior sequences. 
For instance, MBN \cite{mbn} uses a meta multi-behavior sequence encoder to model meta-knowledge across behavior sequences, and a recurring-item-aware predictor to predict duplicated items in the sequences. 
TGT \cite{tgt} utilizes a behavior-aware transformer \cite{transformer} network to capture the short-term interest in multi-behavior sequences, and a temporal graph neural network to capture the multi-behavior dependencies. 
Although they utilize the multi-behavior sequences to capture the sequential patterns of user interests, they cannot achieve optimal performance because they do not fully consider the characteristics of multi-behavior sequences such as data imbalance and heterogeneity.

\subsection{Multi-behavior Recommender Systems}
MBRSs predict the next item of users on a target behavior, modeling the user interests from their multi-behavior data. 
Some methods use the multi-behavior data as auxiliary information about the user interests on the target behavior: 
HMG-CR \cite{hmgcr} uses graph neural networks based on hyper meta-paths between the user's behaviors on an item, and a graph contrastive learning between them. 
CML \cite{cml} uses a contrastive meta network to model cross-type behavior dependencies via self-supervised learning. 
On the other hand, others treat the multi-behavior recommendation as a multi-task problem: 
METAS \cite{metas} designs the user-item relations by dividing them into an action space and an entity space, and learns them with multi-task metric learning. 
EHCF \cite{ehcf} learns the user-item relations through transfer learning between the behaviors, and is trained with an efficient non-sampling method and multi-task learning.

Lastly, some methods \cite{mbn,tgt} use the sequential information in multi-behavior sequences, as mentioned above. 
We note that the sequential information is significant in MBRS, as a user's next item is determined as a result of the drifts of various interests revealed in the multi-behavior sequences.
However, it is difficult to effectively discover the various interests, as the high-level information about the interests is involved in each sequence in complex relationships between them. 
Therefore, we focus on how to effectively capture the user interests in the multi-behavior sequences considering their characteristics.

\subsection{Dynamic Routing}
Dynamic routing was proposed with CapsNet \cite{capsnet} to effectively combine several capsules \cite{capsule}, each of which is a vector neuron in neural networks and encodes various properties of an entity (e.g., an object in an image) in a high-dimensional manner. 
It derives the final result considering the various properties of entities, by dynamically integrating the capsules required in the integration result via iterative routing phases considering the influence of each capsule in the result.
In RSs, the dynamic routing is often used to integrate diverse information about user interests from data. 
CARP \cite{carp} extracts users' sentiments from user reviews by modeling various properties of the reviews with the dynamic routing. 
JTCN \cite{jtcn} uses dynamic routing to estimate high-level user preferences with the side information for cold-start users. 
On the other hand, we employ the dynamic routing to extract personalized information from multi-behavior sequences by dynamically integrating the important capsules encoding heterogeneous information about user interests, which is the first method that applies the dynamic routing to multi-behavior sequence modeling.


\section{Method}

This section explains the proposed methods that model the multi-behavior sequences considering their characteristics. 
We propose two methods: \proposedA, which extracts personalized and heterogeneous information from multi-behavior sequences based on their correlations, and \proposedB, which further models item-level heterogeneity and personalization. 

In this section, let $I, B$ represent the set of indices of items and behaviors, respectively. 
Our proposed methods use a set of each behavior sequence of a user, $S=\bigcup_{b \in B} \{s_b\}$, where $s_b = [i^b_1, i^b_2, ..., i^b_{t_b}]$ and $i^b_k \in I$ is the $k$-th item on which the user took behavior $b$, to predict the next item on the target behavior.
In this paper, we represent a vector as a bold small letter (e.g., $\mathbf{i}^b_k, \mathbf{v}_{d}$), a two-dimensional matrix as a bold capital letter (e.g., $\mathbf{W}^{ir}, \mathbf{H}^{(l)}_k$), and a three-dimensional tensor as a calligraphic capital letter (e.g., $\mathcal{W}^{ir}$).

\subsection{\proposedA}
\begin{figure}[t]
    \centering
    \includegraphics[width=0.9\columnwidth] {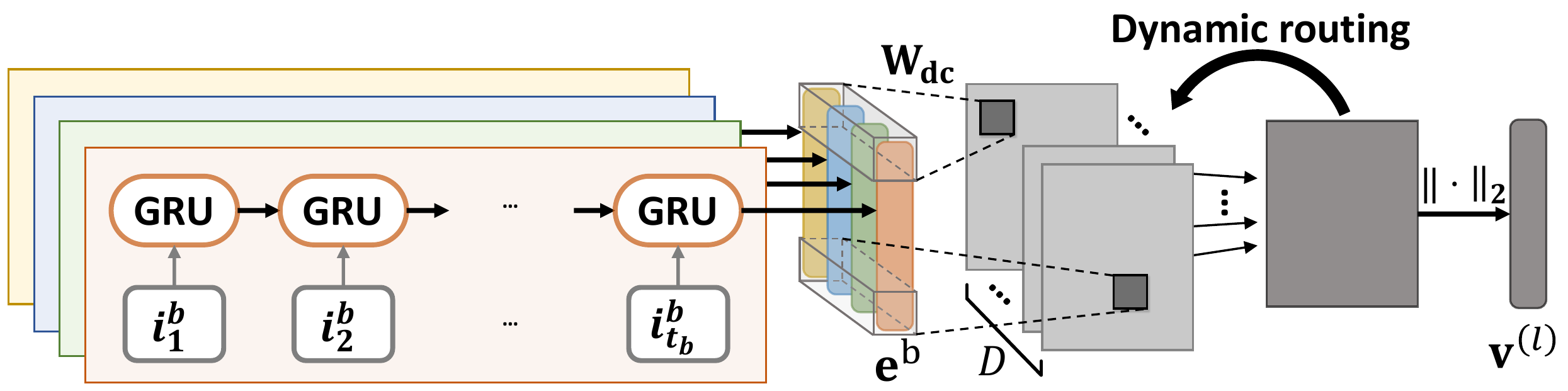}
    \caption{Overall architecture of \proposedA.}
    \label{fig:model1}
\end{figure}

\proposedA (Fig. \ref{fig:model1}) first encodes each behavior sequence of a user with GRU \cite{gru}, which can capture the interest drift of the user within the user-item interaction sequence \cite{gru4rec}, and then dynamically integrates the encoded information through the dynamic routing.

\subsubsection{Sequence Modeling}
For a sequence $s_b = [i^b_1, ..., i^b_{t^b}]$ of behavior $b$, \proposedA first draws a encoded representation $\mathbf{e}^b$ from a GRU for $b$. 
Note that there is a separate GRU for each behavior, although we omit the behavior expression $b$ in the equation for its GRU cell for simplification. 

Specifically, the embedding $\mathbf{i}^b_k \in \mathbb{R}^D$ for the $k$-th item $i^b_k$ of behavior $b$ is encoded in a GRU cell as follows:
\begin{equation}
\fontsize{8}{9.6}\selectfont
    \begin{split}
        \mathbf{r}_k &= \sigma(\mathbf{W}^{ir}\mathbf{i}^b_k + \mathbf{W}^{hr}\mathbf{h}_{k-1} + \mathbf{b}^{r}) \\
        \mathbf{z}_k &= \sigma(\mathbf{W}^{iz}\mathbf{i}^b_k + \mathbf{W}^{hz}\mathbf{h}_{k-1} + \mathbf{b}^{z}) \\
        \mathbf{n}_k &= \tanh(\mathbf{W}^{in}\mathbf{i}^b_k + \mathbf{r}_k * \mathbf{W}^{hn}\mathbf{h}_{k-1} + \mathbf{b}^{n}) \\
        \mathbf{h}_k &= \mathbf{z}_k * \mathbf{n}_k + (\mathbf{1}-\mathbf{z}_k) * \mathbf{h}_{k-1},
    \end{split}
\end{equation}
where $\mathbf{W}^{*} \in \mathbb{R}^{D \times D}$ are weight matrices, $\mathbf{b}^{*} \in \mathbb{R}^{D}$ are biases, $\sigma$ is the sigmoid function and $*$ is the element-wise multiplication.
$\mathbf{h}_k \in \mathbb{R}^D$ is the hidden state for the $k$-th index, which carries the information up to the $k$-th item in the GRU. $\mathbf{h}_0$ is initialized to zeros.
The last hidden state can be the representation which summarizes the overall information within the behavior sequence, that is, $\mathbf{e}^b=\mathbf{h}_{t^b}$. 

\subsubsection{Dynamic Routing}
\proposedA then combines the encoded sequences $\mathbf{e}^b$ for all behaviors $b \in B$ with the dynamic routing.
Dynamic routing \cite{capsnet} dynamically integrates the capsules from among several candidate capsules encoding heterogeneous information about the input entities (e.g., objects in an image), through iterative updates of weight of each candidate capsule based on the integration result. 
Through the iterative updates of the weights considering the result of integrating all information, the model can learn to pay attention to the heterogeneous information important to the final result, and this procedure is optimized for each individual input. 
Judging from these advantages, we think the dynamic routing is suitable for multi-behavior sequence modeling, which needs to obtain \textit{personalized} and \textit{heterogeneous} information from each sequence considering the \textit{correlations} between the sequences. 
We also modify the dynamic routing to be efficient enough for the next item prediction. 
Specifically, \proposedA creates the candidate capsules to be integrated into the final capsules by multiplying a weight matrix to each primary capsule, which is defined as the elements in a certain dimension of each encoded sequence so that each capsule can encode the heterogeneous information from multi-behavior sequences. 
This also involves our modified design for efficient dynamic routing, where the candidate capsules are created for only $D$ final capsules (i.e., number of the final capsules $C$ = $D$), which is much smaller than $|I|$ in general. 
That is, the $c$-th candidate capsule $\mathbf{u}^d_c$ for $c=1,...,C$, generated from the $d$-th primary capsule, is
\begin{equation}
\fontsize{8}{9.6}\selectfont
    \begin{split}
        \mathbf{u}^d_c &= \mathbf{W}_{dc} \times \left[e^1_d \quad ... \quad e^{|B|}_d \right]^\top \in \mathbb{R}^{L}, 
    \end{split}
\end{equation}
where $e^b_d$ is the element in dimension $d$ of $\mathbf{e}^b$, $L$ is the length (i.e., capacity) of a capsule vector, and $\mathbf{W}_{dc} \in \mathbb{R}^{L \times |B|}$ is the weight matrix for the $c$-th candidate capsule for the $d$-th primary capsule: there are $ D^2 \times L \times |B|$ parameters in total. 

Then, the coefficient $c^{(l)}_{dc}$, which determines which of the $C$ candidates capsules generated from the $d$-th primary capsule will be important to the final capsules, is iteratively computed for $l=1,...,r$: 
\begin{equation}
\fontsize{8}{9.6}\selectfont
\label{eq:3}
    \begin{split}
        c^{(l)}_{dc} &= {\exp\left(b^{(l)}_{dc}\right) \over \sum^C_{c'} \exp\left(b^{(l)}_{dc'}\right)}, \\ 
    \end{split}
\end{equation}
where $b^{(l)}_{dc}$ is the logit for coefficient. 
The number of iterations $r$ is a hyperparameter. 
The initial logit $b^{(1)}_{dc}$ is initialized to zero so that each candidate capsule is equally involved in the final capsule, and is iteratively updated to dynamically determine the coefficients when $l > 1$. 

The final capsules $\mathbf{v}^{(l)}_c$, for $c=1,...,C(=D)$, are obtained by integrating the candidate capsules using the coefficients as follows:
\begin{equation}
\fontsize{8}{9.6}\selectfont
    \begin{split}
        \mathbf{v}^{(l)}_c &= \alpha \sum_{d=1}^D c^{(l)}_{dc} \mathbf{u}^d_c \in \mathbb{R}^{C},
    \end{split}
\end{equation}
where $\alpha$ is a learning parameter for scaling.


\begin{figure*}[t]
    \centering
    \includegraphics[width=0.65\textwidth] {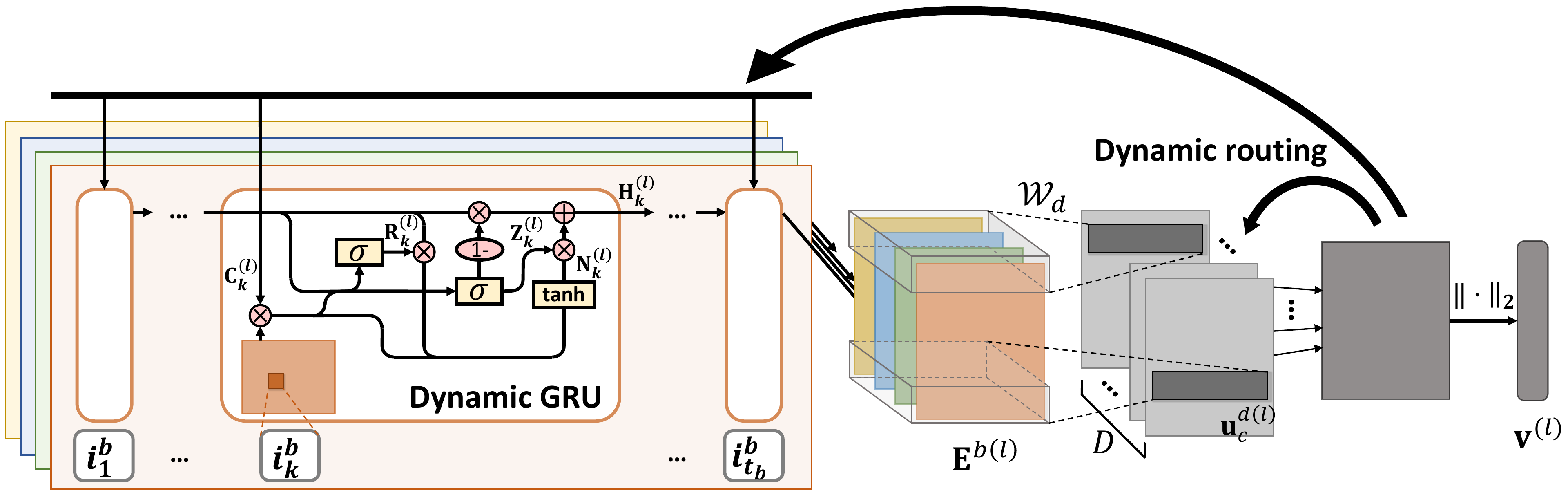}
    \caption{Overall architecture of \proposedB.}
    \label{fig:model2}
\end{figure*}

\proposedA then summarizes the values in each final capsule using its length to sum up the heterogeneous information in the capsule, to obtain the value of a dimension of the final representation $\mathbf{v}^{(l)}$:
\begin{equation}
\fontsize{8}{9.6}\selectfont
    \begin{split}
        \mathbf{\bar{v}}^{(l)} &= \left[\lVert \mathbf{v}^{(l)}_1 \rVert_2 \quad \lVert \mathbf{v}^{(l)}_2 \rVert_2 \quad ... \quad \lVert \mathbf{v}^{(l)}_C \rVert_2 \right]^\top \in \mathbb{R}^{C} \\
        \mathbf{v}^{(l)} &= \mathbf{w} \cdot \mathbf{\bar{v}}^{(l)} + \mathbf{b} \in \mathbb{R}^{C}.
    \end{split}
\end{equation}
As the length of a capsule is positive, a weight vector $\mathbf{w} \in \mathbb{R}^{C}$ and a bias $\mathbf{b} \in \mathbb{R}^{C}$ is applied to the final representation to extend the range of each value to all real numbers. 

When $l < r$, the logit $b^{(l)}_{dc}$ is updated for the subsequent routing phase, reflecting the contribution of each candidate capsule to the integration result, which is represented by a similarity score (e.g., dot product) between them. 
As the integration result, which is a criterion for estimating the contribution, we use both the final capsule and the predicted item embedding $\mathbf{p}^{(l)}$ based on the current final representation:
\begin{equation}
\fontsize{8}{9.6}\selectfont
    \begin{split}
        \mathbf{p}^{(l)} = &\sum_{i \in I} \mathbf{i}_i \cdot \text{softmax}(\mathbf{v}^{(l)} \cdot \mathbf{i}_i) = \sum_{i \in I} \mathbf{i}_i \cdot {\exp(\mathbf{v}^{(l)} \cdot \mathbf{i}_i) \over \underset{i'\in I}{\sum} \exp(\mathbf{v}^{(l)} \cdot \mathbf{i}_{i'})} \\
        \mathbf{r}_c^{(l)} &= \mathbf{W}_{c}^{\text{coef}} \times \left[\mathbf{v}^{(l)}_c \parallel \mathbf{p}^{(l)}\right]^\top\\
        b^{(l+1)}_{dc} &= b^{(l)}_{dc} + \mathbf{u}^d_c \cdot \mathbf{r}_c^{(l)},
    \end{split}
\end{equation}
where $\mathbf{p}^{(l)} \in \mathbb{R}^{D}$ is the predicted item embedding with the $l$-th routing, $\mathbf{r}_c^{(l)} \in \mathbb{R}^{L}$ is the integration result to estimate the contribution for the $c$-th candidate,  $\mathbf{W}_c^{\text{coef}} \in \mathbb{R}^{L \times (C+D)}$ is a weight matrix for the $c$-th candidates, $\parallel$ is the vector concatenation, and $\cdot$ is the dot product. 
If $l<r$, the subsequent routing starts again from Eq. \ref{eq:3} with the updated logits. 
With iterative routings, the weight matrix learns how to give attentions to the candidates important to the final capsule.

After $r$ iterations of routing, \proposedA uses the final representation $\mathbf{v}^{(r)}$ to compute the score of each item as the next item of the user. 
\proposedA is designed efficiently to obtain the score of each item by creating only $D$ final capsules, instead of $|I|$ capsules for solving it as a classification task. 
Given a set of multi-behavior sequences of a user, $S=\bigcup_{b \in B} \{s_b\}$, the estimated probability $\hat{y}_{Si}$ of an item $i \in I$ to be the next item on the target behavior is obtained as follows: 
\begin{equation}
\fontsize{8}{9.6}\selectfont
    \begin{split}
        \hat{y}_{Si} &= \text{softmax}(\mathbf{v}^{(r)} \cdot \mathbf{i}_i) = {\exp(\mathbf{v}^{(r)} \cdot \mathbf{i}_i) \over \sum_{i' \in I} \exp(\mathbf{v}^{(r)} \cdot \mathbf{i}_{i'})}.
    \end{split}
\end{equation}
where $\mathbf{i}_i \in \mathbb{R}^{D}$ is the embedding of item $i$.

\subsection{\proposedB}
\label{section:modelb}

Though \proposedA considers the correlations between multi-behavior sequences via dynamic routing to discover the heterogeneous and personalized information of them, it has a limitation in capturing the heterogeneous and personalized information at \textit{item-level} as it encodes each sequence independently. 
Therefore, we propose \proposedB (Fig. \ref{fig:model2}), with \textit{dynamic GRU} where the dynamic routing is also applied to modeling each sequence to let the correlations affect modeling the heterogeneity and personalization at \textit{item-level}.

\subsubsection{Dynamic Sequence Modeling}
To take the advantages of dynamic routing based on the capsules, a dynamic GRU in \proposedB constructs its internal hidden state as capsules. 
To be specific, the state of each input item is dynamically adjusted based on the final integration result of all information, by the relation between the integration result and each item which is updated through $r$ iterations. 
As a result, the role of each item in a sequence is affected by the other sequences.

Firstly, we define an operation for capsule-wise multiplication using a weight tensor for the capsules in dynamic GRU. 
Given a set of $D$ capsules $\mathbf{H}=[\mathbf{h}_1 \quad ... \quad \mathbf{h}_D]^\top \in \mathbb{R}^{D \times L_1}$ and a weight tensor $\mathcal{W}=[\mathbf{W}_1 \quad ... \quad \mathbf{W}_D]^\top \in \mathbb{R}^{D \times L_2 \times L_1}$, the capsule-wise multiplication $\otimes$ is defined as:
\begin{equation}
\fontsize{8}{9.6}\selectfont
    \begin{split}
        \mathcal{W} \otimes \mathbf{H} &= [\mathbf{W}_1 \mathbf{h}_1 \quad ... \quad \mathbf{W}_D \mathbf{h}_D]^\top \in \mathbb{R}^{D \times L_2}.
    \end{split}
\end{equation}

For each behavior $b \in B$, to make the encoded sequence representation $\mathbf{E}^{b(l)}$ in the $l$-th iteration, our proposed dynamic GRU takes the embedding $\mathbf{i}^b_k \in \mathbb{R}^D$ of the $k$-th item $i^b_k$ to make the next hidden state as follows:
\begin{equation}
\fontsize{8}{9.6}\selectfont
    \begin{split}
        \mathbf{R}^{(l)}_k &= \sigma\left(\mathcal{W}^{ir}\mathbf{i}^b_k * (\mathcal{W}^{cr} \otimes \mathbf{C}^{(l)}_k + \mathbf{B}^{cr}) + \mathcal{W}^{hr} \otimes \mathbf{H}^{(l)}_{(k-1)} + \mathbf{B}^{r}\right) \\
        \mathbf{Z}^{(l)}_k &= \sigma\left(\mathcal{W}^{iz}\mathbf{i}^b_k * (\mathcal{W}^{cz} \otimes \mathbf{C}^{(l)}_k + \mathbf{B}^{cz}) + \mathcal{W}^{hz} \otimes \mathbf{H}^{(l)}_{(k-1)} + \mathbf{B}^{z} \right)\\
        \mathbf{N}^{(l)}_k &= \tanh\left(\mathcal{W}^{in}\mathbf{i}^b_k * (\mathcal{W}^{cn} \otimes \mathbf{C}^{(l)}_k + \mathbf{B}^{cn}) \right.\\
                        & \left. \qquad \qquad \qquad \qquad \qquad \qquad + \mathbf{R}^{(l)}_k * (\mathcal{W}^{hn} \otimes \mathbf{H}^{(l)}_{(k-1)}) + \mathbf{B}^{n} \right)\\
        \mathbf{H}^{(l)}_k &= \mathbf{Z}^{(l)}_k * \mathbf{N}^{(l)}_k + (\mathbf{1}-\mathbf{Z}^{(l)}_k) * \mathbf{H}^{(l)}_{(k-1)},
    \end{split}
\end{equation}
where $\mathbf{H}^{(l)}_k \in \mathbb{R}^{D \times L}$ is the hidden state for the $k$-th index, and $\mathbf{C}^{(l)}_k \in \mathbb{R}^{D \times L}$ is the adjusting state based on the relation between the $k$-th item and the integration result, which is initialized to zeros when $l=1$ and iteratively updated using the integration result.
The weight tensors $\mathcal{W}^{ir}, \mathcal{W}^{iz}, \mathcal{W}^{in} \in \mathbb{R}^{D \times L \times D}$ change the input embedding into the form of $D$ capsules. 
The other weight tensors (i.e., $\mathcal{W}^{*} \in \mathbb{R}^{D \times L \times L}$) and biases (i.e., $\mathbf{B}^{*} \in \mathbb{R}^{D \times L}$) are for the capsules. 


The differences of the dynamic GRU in \proposedB compared to the original GRU in \proposedA are 1) that the hidden state consists of capsules, and 2) that the adjusting state $\mathbf{C}^{(l)}_k$ changes the state of input item over $r$ iterations. 
It can maintain various properties of items within multi-behavior sequences in the capsules from inside the dynamic GRU, and dynamically emphasize the important information considering the correlations with other sequences. 

Similarly to \proposedA, the last hidden state $\mathbf{H}^{(l)}_{t^b}$ is the summarized information of the sequence of behavior $b$ in the $l$-th iteration, that is, $\mathbf{E}^{b(l)}=\mathbf{H}^{(l)}_{t^b} \in \mathbb{R}^{D \times L}$. Using each row of $\mathbf{E}^{b(l)}$ for all behaviors $b \in B$, the candidate capsules $\mathbf{u}^{d(l)}_c$ for $c=1,...,C(=D)$ for the $d$-th dimension are defined as:
\begin{equation}
\fontsize{8}{9.6}\selectfont
\begin{split}
    [\mathbf{u}^{d(l)}_1 ... \ \mathbf{u}^{d(l)}_c ... \ \mathbf{u}^{d(l)}_C] = \mathcal{W}_{d} \otimes \left[\mathbf{e}^{1(l)}_d \ ... \ \ \mathbf{e}^{|B|(l)}_d\right] \in \mathbb{R}^{L \times C}
\end{split}
\end{equation}
where $\mathbf{e}^{b(l)}_d \in \mathbb{R}^{L}$ is the $d$-th row of $\mathbf{E}^{b(l)}$, and $\mathcal{W}_{d} \in \mathbb{R}^{L \times C \times |B|}$ is the weight tensor. 

The other parts for integrating the candidate capsules (i.e., $c^{(l)}_{dc}, \mathbf{v}^{(l)}_c, \mathbf{v}^{(l)}, \mathbf{r}_c^{(l)}, \mathbf{p}^{(l)}$, and $b^{(l+1)}_{dc}$) are similar as in \proposedA.
Finally, for each dynamic GRU, the adjusting state for the $k$-th item for dynamic routing is updated when $l < r$. 
To maintain the adjusting state in the form of capsules, it is updated with element-wise multiplication between the integration result and $\mathbf{N}^{(l)}_k$, which is the new gate in the dynamic GRU encoding the information about the input item:
\begin{equation}
\fontsize{8}{9.6}\selectfont
    \begin{split}
        \mathbf{C}^{(l+1)}_k &= \mathbf{C}^{(l)}_k + \mathbf{N}^{(l)}_k * [\mathbf{r}^{(l)}_1 \quad ... \quad \mathbf{r}^{(l)}_C] \in \mathbb{R}^{D \times C}.
    \end{split}
\end{equation}
The adjusting state $\mathbf{C}^{(l+1)}_k$ learns to adjust the input state in each dynamic GRU in the next iteration to encode important information in the capsules, with the relation between the integration result in the $l$-th iteration and the $k$-th item on behavior $b$. 
Finally, the estimated probability $\hat{y}_{Si}$ are also defined similarly to \proposedA.


\subsection{Prediction}
We use the binary cross-entropy loss as the loss function $L$:
\begin{equation}
\fontsize{7.8}{9.36}\selectfont
    \begin{split}
        L &= - {1 \over |S^T|} \sum_{S \in S^T} \sum_{i \in I} \left(y_{Si}\log \hat{y}_{Si} + (1-y_{Si})\log(1 - \hat{y}_{Si})\right),
    \end{split}
\end{equation}
where $S^T$ is the sets of a user's multi-behavior sequences in the training set, and $y_{Si} \in \{0, 1\}$ is the ground-truth label which indicates whether item $i$ is the true next item of the set of multi-behavior sequences $S$.


\section{Experiments}
\subsection{Experimental Settings}

\subsubsection{Dataset}
Our experiments were conducted on two public datasets: \textit{Taobao}\footnote{https://tianchi.aliyun.com/dataset/dataDetail?dataId=649} and \textit{Tmall}\footnote{https://tianchi.aliyun.com/dataset/dataDetail?dataId=47}, and a new dataset we release: \textit{GS Retail}\footnote{http://di.postech.ac.kr}. 
Table \ref{table:statistics} provides the statistics of the datasets. 
For all datasets, we filtered out users and items that have less than five interactions on the target behavior (i.e., Purchase), and used recent 500 interactions of each user. 

\textit{GS Retail} is a new dataset we release. User behavior data of GS SHOP, which is an e-commerce and home-shopping brand of GS Retail, was collected for a month in 2021. This dataset contains three types of user behaviors: Purchase, Add-to-cart, and Click. Compared with the other datasets, this dataset has the largest and the most up-to-date data. 


\begin{table}[t]
\centering
\fontsize{9}{10.8}\selectfont
\begin{tabular}{l|ccc}
\hline
                  & Taobao                 & Tmall                      & GS Retail              \\ \hline
\#Users            & 987,994                & 424,170                   & 13,334,687 \\
\#Items            & 4,162,024              & 1,090,390                 & 3,477,502 \\
\#Purchases        & 2,015,839              & 3,292,144                 & 6,510,474 \\
\#Favorites        & 2,888,258              & 3,005,723                 & -         \\
\#Add-to-carts     & 5,530,446              & 76,750                    & 15,283,350 \\
\#Clicks           & 89,716,264             & 48,550,713                & 158,567,115 \\
Time period        & 9 days                 & 186 days                   & 32 days                  \\
Year               & 2017                   & $\le$2014                  & 2021                     \\ \hline
\end{tabular}
\caption{Statistics of datasets.} 
\label{table:statistics}
\end{table}

\begin{table*}[t]
\centering
\fontsize{9}{10.8}\selectfont
\setlength{\tabcolsep}{1.5pt}
\centering
\begin{tabular}{c|c|cccccccccccc|c}
\hline
                            Dataset &  Metric & {BPR-MF}    & {GRU4Rec} & {SASRec}       & \multicolumn{1}{c|}{{FMLP-Rec}}         & {METAS}  & {EHCF}   & {HMG-CR} & {CML}    & {MBN}          & \multicolumn{1}{c|}{{TGT}}          & {\proposedA} & {\proposedB} & $Imp.$(\%)      \\ \hline
\multirow{6}{*}{\rotatebox[origin=c]{90}{Taobao}}     & H@10  & 0.0105 & 0.0945  & 0.1391       & \multicolumn{1}{c|}{\underline{0.2286}}       & 0.0062 & 0.0315 & 0.0023 & 0.0017 & 0.2089       & \multicolumn{1}{c|}{0.1768} & 0.2758$^*$    & \textbf{0.3166}$^*$ & 38.5\% \\
                            & H@20  & 0.0151 & 0.1095  & 0.1565       & \multicolumn{1}{c|}{\underline{0.2509}}       & 0.0097 & 0.0425 & 0.0041 & 0.0026 & 0.2259       & \multicolumn{1}{c|}{0.2147} & 0.2947$^*$    & \textbf{0.3358}$^*$ & 33.8\% \\
                            & N@10  & 0.0058 & 0.0659  & 0.1030       & \multicolumn{1}{c|}{\underline{0.1772}}       & 0.0034 & 0.0184 & 0.0013 & 0.0009 & 0.1630 & \multicolumn{1}{c|}{0.1239}       & 0.2101$^*$    & \textbf{0.2369}$^*$ & 33.7\% \\
                            & N@20  & 0.0070 & 0.0696  & 0.1074       & \multicolumn{1}{c|}{\underline{0.1828}}       & 0.0043 & 0.0212 & 0.0017 & 0.0012 & 0.1674 & \multicolumn{1}{c|}{0.1335}       & 0.2149$^*$    & \textbf{0.2418}$^*$ & 32.3\% \\ \cline{2-15} 
                            & \#Prm &    89.1M& 153.4M&  19.1M        & \multicolumn{1}{c|}{76.6M}       &  44.6M& 37.1M&    11.2M&  22.3M &  89.4M & \multicolumn{1}{c|}{77.4M}             &     77.4M   &     81.7M & - \\
                            & Inf.t &    17.2 & 15.6  &  14.2        & \multicolumn{1}{c|}{15.8}         & 20.8   & 25.0     & 20.2 & 17.4   &  32.5 & \multicolumn{1}{c|}{23.5}             & 20.4        & 28.6      & -       \\ 
                            \hline
\multirow{6}{*}{\rotatebox[origin=c]{90}{Tmall}}      & H@10  & 0.0092 & 0.0554  & 0.0720       & \multicolumn{1}{c|}{0.0862} & 0.0049 & 0.0319 & 0.0052 & 0.0028 & \underline{0.0980}       & \multicolumn{1}{c|}{0.0331}       & 0.0982    & \textbf{0.1380}$^*$ & 40.8\% \\
                            & H@20  & 0.0160 & 0.0689  & 0.0789       & \multicolumn{1}{c|}{0.1026} & 0.0086 & 0.0428 & 0.0084 & 0.0040 & \underline{0.1189}       & \multicolumn{1}{c|}{0.0465}       & 0.1152    & \textbf{0.1576}$^*$ & 32.6\% \\
                            & N@10  & 0.0044 & 0.0380  & 0.0608 & \multicolumn{1}{c|}{0.0612}       & 0.0023 & 0.0186 & 0.0021 & 0.0013 & \underline{0.0630}       & \multicolumn{1}{c|}{0.0204}       & 0.0694$^*$    & \textbf{0.0941}$^*$ & 49.4\% \\
                            & N@20  & 0.0061 & 0.0413  & 0.0625 & \multicolumn{1}{c|}{0.0654}       & 0.0033 & 0.0203 & 0.0029 & 0.0018 & \underline{0.0692}       & \multicolumn{1}{c|}{0.0237}       & 0.0737$^*$    & \textbf{0.0990}$^*$ & 43.1\% \\ \cline{2-15} 
                            & \#Prm & 18.5M   & 23.0M & 2.9M  & \multicolumn{1}{c|}{5.8M}             & 18.6M     & 37.1M & 4.7M  & 9.3M  & 37.3M & \multicolumn{1}{c|}{5.7M}             & 12.5M     &  13.9M & - \\
                            & Inf.t & 4.3    &    5.7 & 5.0   & \multicolumn{1}{c|}{6.2 }             & 6.7      & 25.7  & 19.4  & 8.8   & 23.1  & \multicolumn{1}{c|}{11.4}             & 10.6        & 18.0     & -          \\ 
                            \hline
\multirow{6}{*}{{\rotatebox[origin=c]{90}{GS Retail}}} & H@10 & 0.0315 & 0.4150  & 0.4119 & \multicolumn{1}{c|}{0.4061}       & 0.0248 & 0.0649 & 0.0241 & 0.0038 & \underline{0.4345}       & \multicolumn{1}{c|}{0.2661}       & 0.4988$^*$    & \textbf{0.5409}$^*$ & 24.5\% \\
                            & H@20  & 0.0485 & 0.4341  & 0.4152 & \multicolumn{1}{c|}{0.4470}       & 0.0427 & 0.0842 & 0.0364 & 0.0067 & \underline{0.4694}       & \multicolumn{1}{c|}{0.3135}       & 0.5287$^*$    & \textbf{0.5692}$^*$ & 21.3\% \\
                            & N@10  & 0.0164 & 0.3774  & \underline{0.4036} & \multicolumn{1}{c|}{0.3253}       & 0.0118 & 0.0402 & 0.0125 & 0.0018 & 0.3443       & \multicolumn{1}{c|}{0.1948}       & 0.4103$^*$    & \textbf{0.4270}$^*$ & 5.8\% \\
                            & N@20  & 0.0207 & 0.3822  & \underline{0.4044} & \multicolumn{1}{c|}{0.3356}       & 0.0163 & 0.0451 & 0.0156 & 0.0024 & 0.3531       & \multicolumn{1}{c|}{0.2068}       & 0.4179$^*$    & \textbf{0.4342}$^*$ & 7.4\% \\ \cline{2-15} 
                            & \#Prm & 42.3M  & 20.5M   & 5.1M & \multicolumn{1}{c|}{10.4M}             &   21.2M     &   42.4M    & 5.3M   & 10.6M   &   42.5M     & \multicolumn{1}{c|}{5.1M}         & 10.7M     & 12.0M & - \\
                            & Inf.t & 7.2    & 5.1     & 5.5  & \multicolumn{1}{c|}{7.1}             & 8.1       & 27.4     &    60.3    & 7.6      &    32.6      & \multicolumn{1}{c|}{12.5}         & 10.8      & 19.4     & -          \\ 
                            \hline
\end{tabular}
\caption{Overall performance of multi-behavior-based methods on the next item prediction. For each dataset and metric, the best performance is highlighted in boldface, and the best performance among the competitors is underlined. $Imp.$ is the improvement of \proposedB over the best competitor. \#Prm is the number of parameters in the model, and Inf.t is the inference time (in seconds). * indicates the improvement over the best competitor is statistically significant with $p<0.01$, using the student $t$-test.}
\label{table:main}
\end{table*}

\begin{table}[t]
\centering
\fontsize{9}{10.8}\selectfont
\setlength{\tabcolsep}{1.3pt}
\begin{tabular}{c|c|cccc}
\hline
Dataset                     & Metric & BPR-MF & GRU4Rec & SASRec & FMLP-Rec \\ \hline
\multirow{6}{*}{\rotatebox[origin=c]{90}{Taobao}}     & H@10   & 0.0414 & 0.0190  & 0.0409 & 0.0464   \\
                            & H@20   & 0.0487 & 0.0225  & 0.0431 & 0.0489   \\
                            & N@10   & 0.0264 & 0.0136  & 0.0347 & 0.0412   \\
                            & N@20   & 0.0282 & 0.0144  & 0.0352 & 0.0418   \\ \cline{2-6} 
                            & \#Prm  & 89.1M  & 153.4M  & 19.1M  & 38.2M    \\
                            & Inf.t  & 17.4   & 15.8    & 14.1   & 15.6     \\ \hline
\multirow{6}{*}{\rotatebox[origin=c]{90}{Tmall}}      & H@10   & 0.0101 & 0.0067  & 0.0093 & 0.0092   \\
                            & H@20   & 0.0170 & 0.0104  & 0.0130 & 0.0131   \\
                            & N@10   & 0.0050 & 0.0039  & 0.0064 & 0.0062   \\
                            & N@20   & 0.0068 & 0.0048  & 0.0073 & 0.0072   \\ \cline{2-6} 
                            & \#Prm  & 37.1M  & 23.0M   & 5.7M   & 5.8M     \\
                            & Inf.t  & 4.3    & 5.6     & 5.1    & 6.0      \\ \hline
\multirow{6}{*}{\rotatebox[origin=c]{90}{GS Retail}} & H@10   & 0.0329 & 0.1028  & 0.1225 & 0.1081   \\
                            & H@20   & 0.0485 & 0.1247  & 0.1399 & 0.1322   \\
                            & N@10   & 0.0174 & 0.0741  & 0.0992 & 0.0766   \\
                            & N@20   & 0.0214 & 0.0796  & 0.1036 & 0.0827   \\ \cline{2-6} 
                            & \#Prm  & 42.3M  & 20.4M   & 2.5M   & 5.1M     \\
                            & Inf.t  & 7.0    & 5.2     & 5.5    & 6.9      \\ \hline
\end{tabular}
\caption{Performance of single-behavior-based methods.}
\label{table:single}
\end{table}

\subsubsection{Evaluation}
We used the most recent interacted item on the target behavior of each user as the test item, and the second most recent one as the validation item, and trained the model with the rest of the data. 
For accurate performance comparison, we ranked the test item by comparing it with \textit{all} other negative items. 
The ranking metrics are Hit Ratio@$n$ (H@$n$) and NDCG@$n$ (N@$n$) which are commonly used for top-$n$ recommendation \cite{timelyrec, concf}, and we used 10, 20 for $n$.

\subsubsection{Baselines}

We compared \proposedA and \proposedB with several types of baselines.
We adopted the single-behavior-based methods, which use only data on the target behavior: 
\textbf{BPR-MF} \cite{bpr} is a matrix factorization method, with a pair-wise loss for personalized ranking. 
\textbf{GRU4Rec} \cite{gru4rec} and \textbf{SASRec} \cite{sasrec} learns the user interests in item sequences, with GRU and a self-attentive network, respectively. 
\textbf{FMLP-Rec} \cite{fmlp}, which is one of the state-of-the-art SRSs, utilizes an all-MLP structure and filtering algorithms to denoise the item sequence. 
We also revised these methods into straightforward multi-behavior versions, in which a unified sequence of the multi-behavior data that consists of the sum of item embedding and behavior embedding is used. 

We also adopted the multi-behavior-based methods:
\textbf{METAS} \cite{metas} designs the relationships between users and items in an action space and an entity space, and learns them with multi-task metric learning. 
\textbf{EHCF} \cite{ehcf} learns user-item relationships with transfer learning between the behaviors, a non-sampling optimization and multi-task learning. 
\textbf{HMG-CR} \cite{hmgcr} adopts graph neural networks based on hyper meta-paths between a user's behaviors, and a graph contrastive learning. 
\textbf{CML} \cite{cml} uses contrastive meta network to model the cross-type behavior dependencies. 
\textbf{MBN} \cite{mbn} is a multi-behavior SRS with an RNN-based meta multi-behavior sequence encoder to capture meta-knowledge between the sequences and a recurring-item-aware predictor. 
\textbf{TGT} \cite{tgt} is a multi-behavior SRS that captures short-term user interests with a behavior-aware transformer network and long-term user interests via a temporal graph neural network. 

\subsubsection{Implementation Details}
We used Adam optimizer \cite{adam} to optimize all models, and tuned the hyperparameters based on the validation N@10 performance: learning rate $\eta \in \{$1e-02, 3e-03, 1e-03, 3e-04, 1e-04$\}$, dropout rate $p \in \{0.0, 0.1, 0.2, 0.3, 0.4, 0.5\}$, coefficient for L2 regularization $\lambda \in \{0.0, 0.001, 0.01, 0.1\}$, embedding dimension $D \in \{16, 32, 64, 128\}$, batch size $\mathcal{B} \in \{64, 128, 256, 512\}$. 
For SRSs, we tuned the length of sequence  $\mathcal{L} \in \{10, 20, 50, 100\}$. 
For \proposedA and \proposedB, the capsule length $L$ is tuned in $\{2, 4, 8, 16, 32\}$ and the number of iterations for dynamic routing $r=2$. 

\subsection{Performance Comparison}
Table \ref{table:main} and \ref{table:single} shows the overall performance of the multi-behavior-based and the single-behavior-based methods, respectively. 
According to the results, firstly, \proposedB showed the best performance on all datasets, and \proposedA was second in most of the results (except for H@20 on Tmall). 
This verifies the superiority of \proposedA and \proposedB to model the user interests within multi-behavior sequences considering the characteristics of them. 
Also, \proposedB, using the dynamic GRUs to capture at item-level heterogeneity and personalization, outperformed \proposedA, which shows the significance of modeling the heterogeneous and personalized information both at sequence- and item-level based on the correlations between the behavior sequences. 
Moreover, our efficient dynamic routing makes our methods have superior performance without too many parameters or much inference time compared to other methods.

Generally, the methods using multi-behavior data, even in a simple manner, have better performance than using only single-behavior data, which means that it is important to exploit the various information of multi-behavior data. 
However, the methods that adopt multi-task learning strategy, especially BPR-MF for multi-behavior and METAS, perform worse than the methods using single-behavior as the models overfit in predicting the items on the most behavior (i.e., Click) rather than the target behavior. 
Also, the methods that model the sequential information provide more accurate predictions than the methods that do not, and the differences are more evident in the multi-behavior scenario. 
These results support our claim that it is important to consider the sequential information, especially in multi-behavior data.

\subsection{Analyses on \proposedA and \proposedB} 
\subsubsection{Hyperparameter Study}
\begin{figure}[t]
    \centering
    \begin{subfigure}[b]{0.4\columnwidth}
        \centering
        \includegraphics[width=\textwidth] {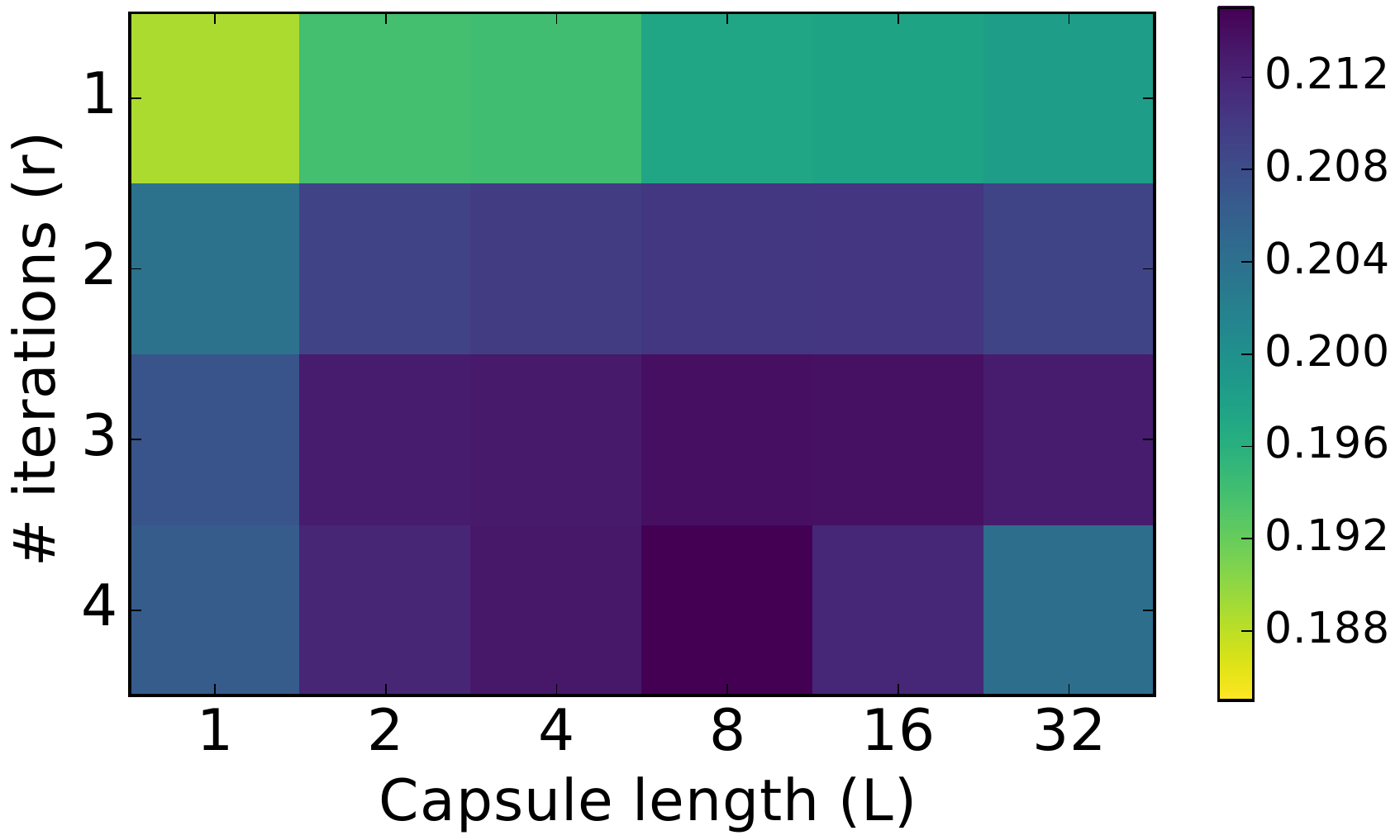}
        \caption{\proposedA}
    \end{subfigure}
    \begin{subfigure}[b]{0.4\columnwidth}
        \centering
        \includegraphics[width=\textwidth] {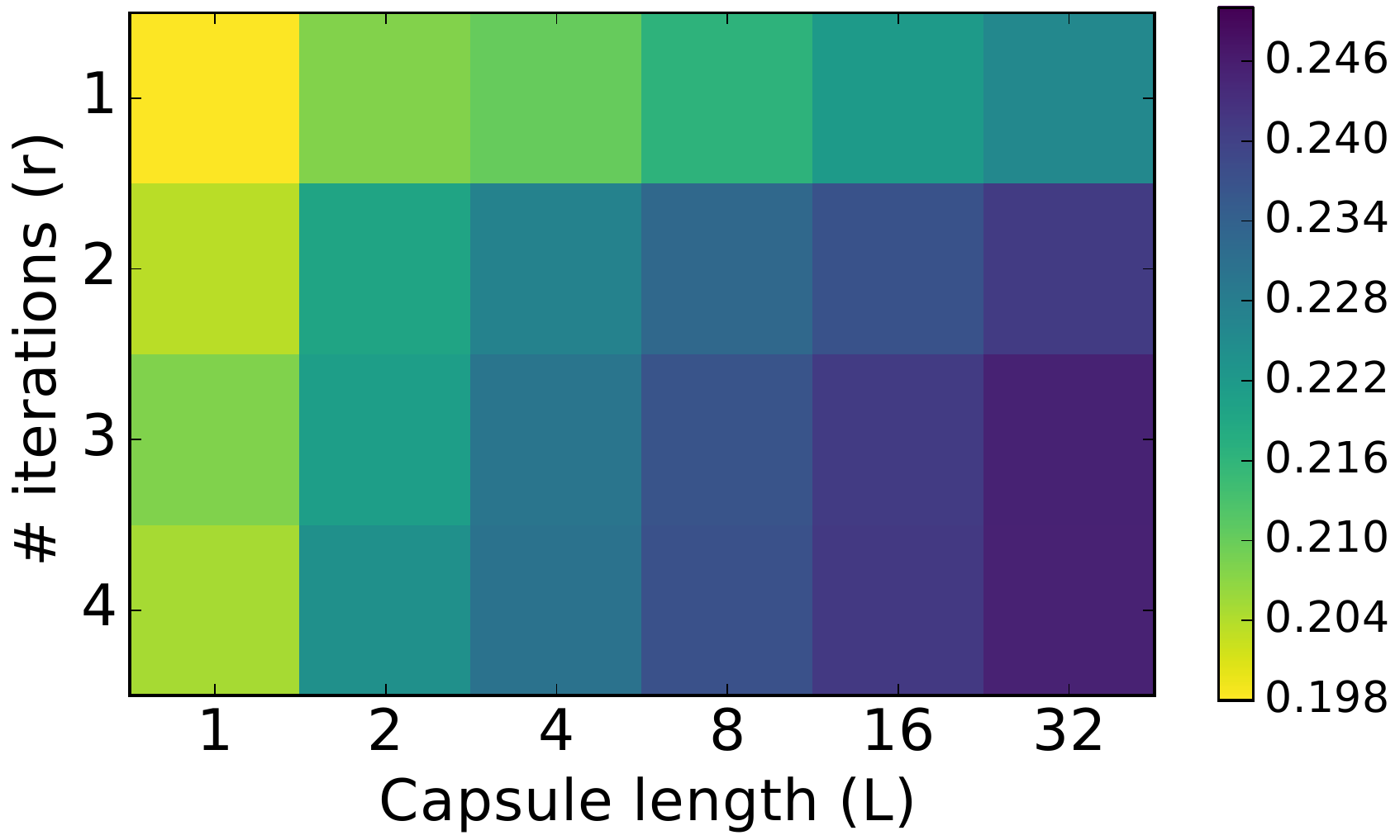}
        \caption{\proposedB}
    \end{subfigure}
    \caption{NDCG@10 performance with various combinations of the capsule length and the number of iterations.}
    \label{fig:hyperparameter}
\end{figure}

We analyse the effects of hyperparameters of \proposedA and \proposedB: the number of iterations for dynamic routing $r$ and the capsule length $L$. 
Fig. \ref{fig:hyperparameter} demonstrates the performance of \proposedA and \proposedB with various combinations of $r$ and $l$ on Taobao. 
Note that the results on the other datasets showed similar trends. 
On both methods, the performance increases as $r$ increases to some extent, which shows the effectiveness of the dynamic routing which makes the model pay attention to the important capsules. 
When $r>2$, the model can refer to more accurate integration results after the dynamic routing in the earlier phases, which leads a further improvement of performance. 
Also, a sufficiently large $L$ increases the performance by providing enough space for the capsules to encode heterogeneous information of the multi-behavior sequences. 
Finally, the optimal capsule length is larger in \proposedB than in \proposedA, which implies the dynamic GRUs can encode more informative and heterogeneous information via its capsule-based structure. 
However, the model overfits and the performance decreases if $r$ or $L$ becomes too large, which implies it is important to select $r$ and $L$ properly.


\subsubsection{Analysis on Dynamic Routing}
\begin{figure}[t]
    \centering
    \includegraphics[width=0.65\columnwidth]{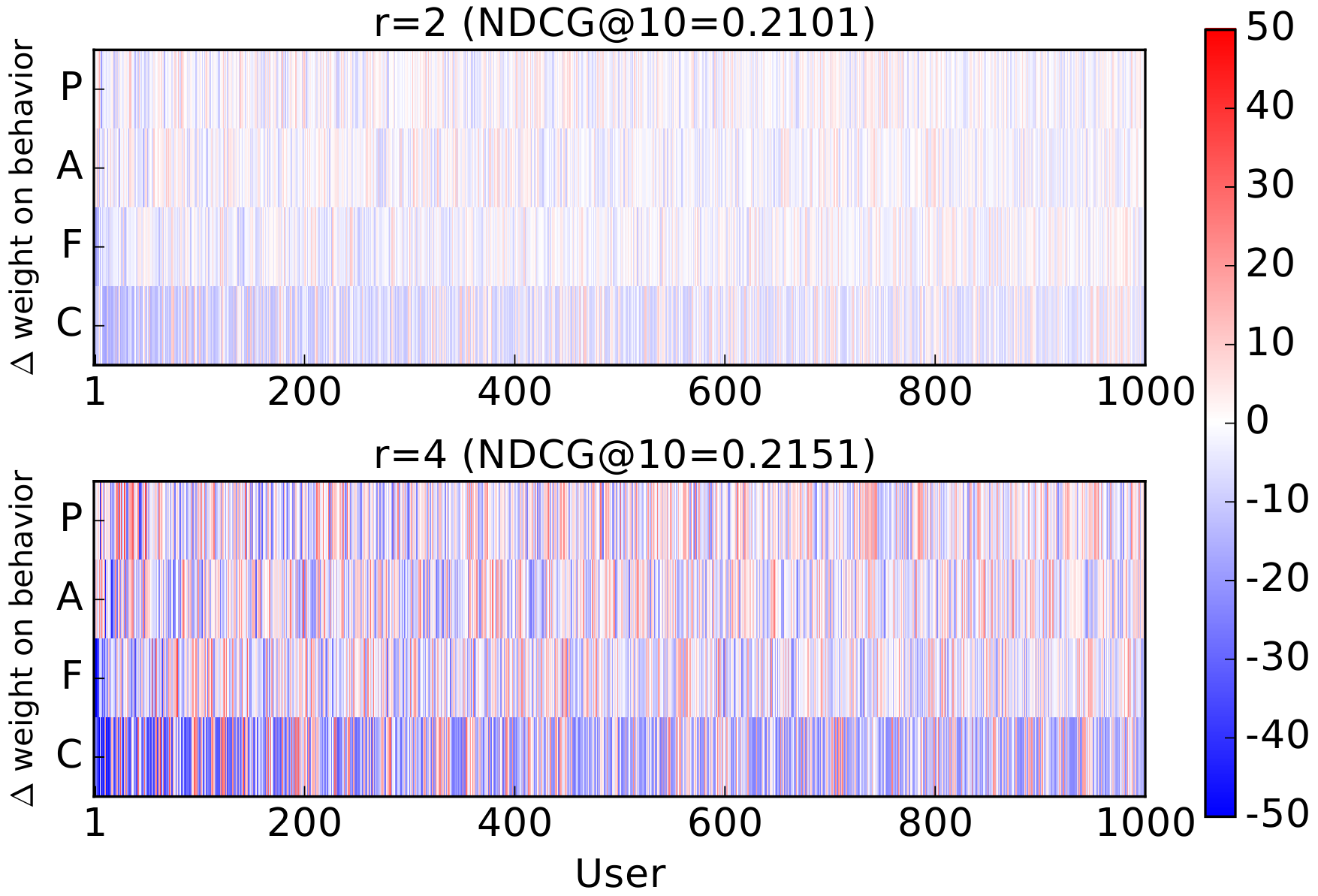}
    \caption{Rates of change of the weights on each behavior compared to the first iteration in \proposedA when $r=2$ and $r=4$ for 1,000 users with the largest total rate of change. P: Purchase, A: Add-to-cart, F: Favorite, C: Click.}
    \label{fig:dymus}
\end{figure}


\begin{table}[t]
\centering
\fontsize{9}{10.8}\selectfont
\setlength{\tabcolsep}{1.3pt}
\begin{tabular}{c|c|cc|cccc}
\hline
            & Original & Sum    & Self-att & -P     & -A     & -F     & -C     \\ \hline
\proposedA  & 0.2101   & 0.1805 & 0.1820   & 0.1212 & 0.2078 & 0.1947 & 0.0872 \\
\proposedB  & 0.2369   & 0.2057 & 0.2187   & 0.1643 & 0.2163 & 0.2011 & 0.1027 \\
MBN         & 0.1630   & -      & -        & 0.1622 & 0.1627 & 0.1631 & 0.0535 \\ \hline

\end{tabular}
\caption{Ablation study on Taobao (N@10). P: Purchase, A: Add-to-cart, F: Favorite, C: Click.}
\label{table:ablation}
\end{table}

To show the dynamic routing in \proposedA effectively pays attention to specific information for each user to obtain personalized information, we show the weights on each behavior varying over iterations through the dynamic routing.
Each value in a column in Fig. \ref{fig:dymus} represents the rate of change of the weight applied to each behavior type in the last iteration of the dynamic routing compared to the first iteration for a user, and a figure reports the 1,000 users with the largest total rate of change.
In other words, each value represents the rate of change of the weight $\mathbf{W}_{dc}$ for the first row of the final capsules (i.e., $\mathbf{v}_1^{(l)}$), which varies according to candidate capsules $\mathbf{u}^d_c$ emphasized by the coefficient $c_{dc}^{(l)}$ updated through the routing iterations. 
The results tell that for both cases, there is a personalized change in the weights on behaviors for each user compared to the first iteration, and the weight change with $r=4$ is greater than that with $r=2$, with a higher performance. 
Note that compared to when $r=1$ $(N@10=0.1971)$ where all candidate capsules are integrated equally without the dynamic routing, the only difference with $r=2$ $(N@10=0.2101)$ and $r=4$ $(N@10=0.2151)$ is that the coefficients can pay attention to specific capsules through the dynamic routing.
Therefore, the results indicate that to train \proposedA to pay attention to specific capsules to obtain the personalized information for each user through the dynamic routing improves performance, and the weights can be personalized more effectively with sufficient iterations. 


\subsubsection{Ablation Study} 
To show that the dynamic routing in \proposedA and \proposedB is more effective than other integration methods, and our methods effectively encode heterogeneous information of the multi-behavior sequences, we performed ablation studies on them.
The results on Taobao are reported in Table \ref{table:ablation}. 
Note that the results on the other datasets showed similar trends. 
Firstly, when the outputs from the behavior GRUs are integrated using sum or self-attention instead of the dynamic routing, the performance decreases because they cannot sufficiently encode the heterogeneous information or personalize the information from the encoded sequences. 
Also, the performance decreases when the influence of each behavior data is removed. 
This phenomenon is more evident on \proposedA and \proposedB than MBN \cite{mbn} which is a state-of-the-art SRS for multi-behavior, which shows the ability of our methods to extract meaningful information from each behavior sequence.

\section{Conclusion}
In this paper, we summarize the characteristics of multi-behavior sequences that have to be considered in SRSs, and propose two novel methods for modeling multi-behavior sequences, \proposedA and \proposedB, which consider the characteristics thoroughly. 
\proposedA adopts the dynamic routing to consider the correlations between the behavior sequences to model the heterogeneous and personalized information, and \proposedB extends the dynamic routing to dynamic GRUs to model them even at item-level.
We also release a new, large and up-to-date dataset for MBRS, which can contribute to the future MBRS studies. 
Our experiments show the superiority of \proposedA and \proposedB compared with the several state-of-the-art baselines, and our analyses on the behaviors of the proposed methods verify the importance of modeling the addressed characteristics of multi-behavior sequences, and the abilities of the proposed methods to model them. 


\section*{Acknowledgements}
This work was supported by Institute of Information $\&$ communications Technology Planning $\&$ Evaluation (IITP) grant funded by the Korea government (MSIT) (No.2018-0-00584, (SW starlab) Development of Decision Support System Software based on Next-Generation Machine Learning), the NRF grant funded by the MSIT (South Korea, No.2020R1A2B5B03097210), and Institute of Information $\&$ communications Technology Planning $\&$ Evaluation (IITP) grant funded by the Korea government (MSIT) (No.2019-0-01906, Artificial Intelligence Graduate School Program(POSTECH))

\bibliography{aaai23}


\end{document}